\documentclass[aps,twocolumn]{revtex4}

\usepackage{amssymb}
\usepackage{amsmath}
\usepackage{bm}
\usepackage[dvips]{epsfig}
\usepackage{graphicx}

\begin{document}

\title{Oscillation phenomena and experimental determination 
of exact mathematical Stability Zones for magneto-conductivity 
in metals having complicated Fermi surfaces.}

\author{A.Ya. Maltsev.}

\affiliation{
\centerline{\it L.D. Landau Institute for Theoretical Physics}
\centerline{\it 142432 Chernogolovka, pr. Ak. Semenova 1A,
maltsev@itp.ac.ru}}

\begin{abstract}
We consider the problem of exact experimental determination
of the boundaries of Stability Zones for magneto-conductivity
in normal metals in the space of directions of $\, {\bf B} \, $.
As can be shown, this problem turns out to be nontrivial since
the exact boundaries of Stability Zones are in fact unobservable
in direct measurements of conductivity. However, this problem
can be effectively solved with the aid of the study of 
oscillation phenomena (cyclotron resonance, quantum oscillations)
in normal metals, which reveal a singular behavior on the
mathematical boundary of a Stability Zone. 
\end{abstract}

\maketitle

\vspace{5mm}

\section{Introduction.}

 In this paper we consider galvano-magnetic phenomena in normal metals,
having complicated Fermi surfaces, in the limit of the strong magnetic 
fields $\, {\bf B} \,\, $ ($ \omega_{B} \tau \, \gg \, 1 $).
In the standard approach we assume that the electron states are
described by a single-particle partition function $\, f ({\bf p}) \, $
defined on the space of the quasimomenta 
$\, {\bf p} \, \in \, \mathbb{T}^{3} \, $ for a given type of
the crystal lattice of a metal. In the equilibrium state we assume
as usually that all the electron states with energy
$\, \epsilon ({\bf p}) \, $ less than the Fermi energy are occupied
($f ({\bf p}) = 1$) and all the states with energy higher than the
Fermi energy are empty ($f ({\bf p}) =	0$). Let us say also that we
will omit here the spin variables which will not play an essential 
role in our consideration.

 The space of the quasimomenta $\, {\bf p} \, $ for a given 
conduction band represents a three-dimensional torus
$\, \mathbb{T}^{3} \, $ given by the factorization of the space
$\, \mathbb{R}^{3} \, $ over the reciprocal lattice vectors
\begin{multline*}
\mathbb{T}^{3} \,\,\,\, = \,\,\,\, \left. \mathbb{R}^{3} \, \right/
\, \left\{ n_{1} {\bf a}_{1} \, + \, n_{2} {\bf a}_{2} \, + \,
n_{3} \, {\bf a}_{3} \right\} \quad ,   \\
n_{1} \, , \,\, n_{2} \, , \,\, n_{3} \,\,\, \in \,\,\, \mathbb{Z}
\end{multline*}

 The vectors 
$\, ({\bf a}_{1}, \, {\bf a}_{2}, \, {\bf a}_{3} ) \, $
represent a basis for the reciprocal lattice and are connected
with the basis vectors
$\, ({\bf l}_{1}, \, {\bf l}_{2}, \, {\bf l}_{3} ) \, $
of the direct lattice by standard relations
$${\bf a}_{1} \,\,\, = \,\,\, 2 \pi \hbar \,\,
{{\bf l}_{2} \, \times \, {\bf l}_{3} \over
({\bf l}_{1}, \, {\bf l}_{2}, \, {\bf l}_{3} )} \,\,\, , \quad \quad
{\bf a}_{2} \,\,\, = \,\,\, 2 \pi \hbar \,\,
{{\bf l}_{3} \, \times \, {\bf l}_{1} \over
({\bf l}_{1}, \, {\bf l}_{2}, \, {\bf l}_{3} )} \,\,\, ,  $$
$${\bf a}_{3} \,\,\, = \,\,\, 2 \pi \hbar \,\,
{{\bf l}_{1} \, \times \, {\bf l}_{2} \over
({\bf l}_{1}, \, {\bf l}_{2}, \, {\bf l}_{3} )} $$

 In this approach the Fermi surface $\, S_{F} \, $:
$$S_{F} \quad : \quad  \epsilon ({\bf p}) \, = \, \epsilon_{F} $$
represents a compact smooth surface embedded into three-dimensional
torus $\, \mathbb{T}^{3} \, $.

\begin{figure}[t]
\begin{center}
\includegraphics[width=0.9\linewidth]{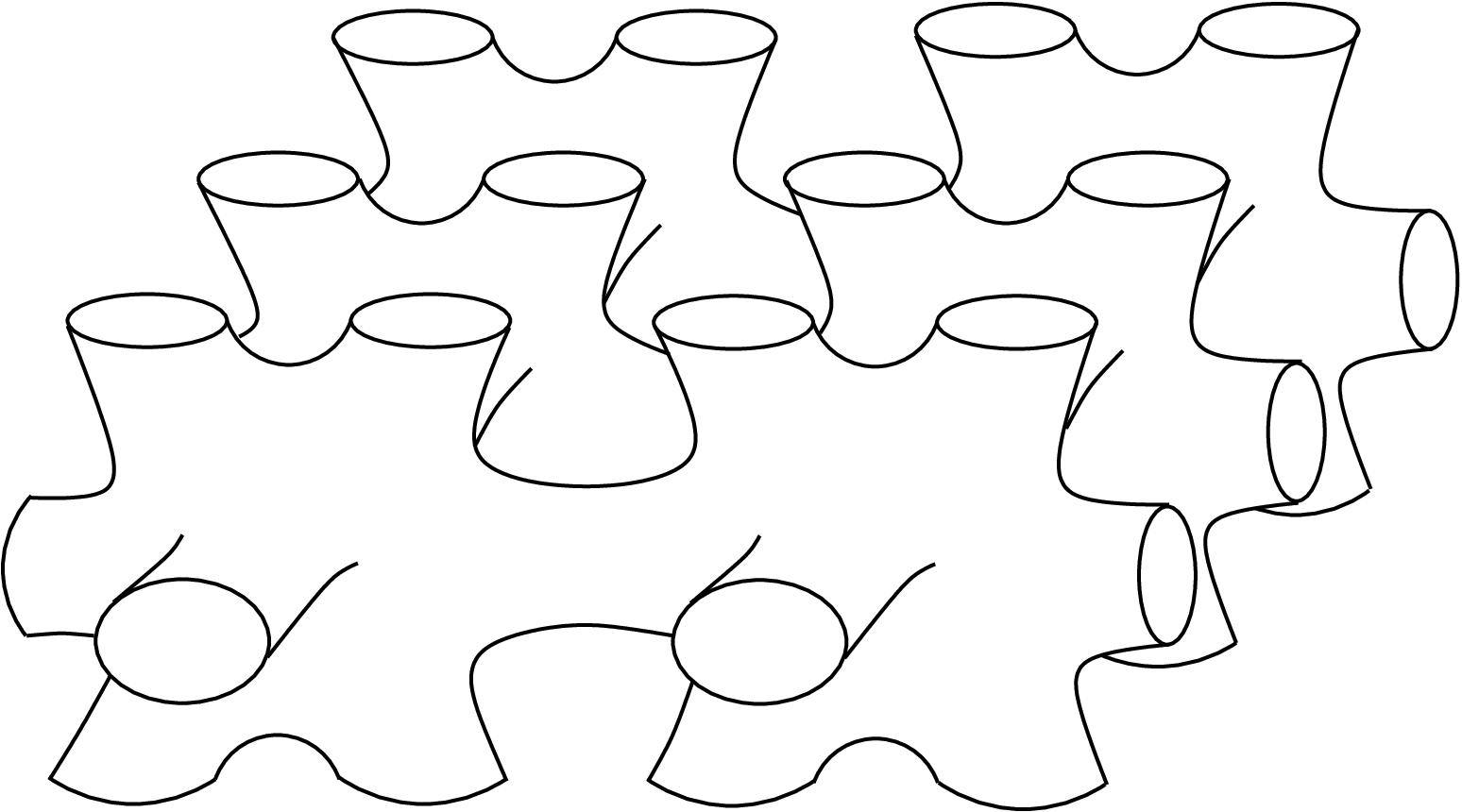}
\end{center}
\caption{A complicated Fermi surface as a 3-periodic surface in the 
${\bf p}$ - space.}
\label{FermiSurf}
\end{figure}

 Equivalently, we can consider the whole space of the quasimomenta
($\mathbb{R}^{3}$) and assume that any two values of $\, {\bf p} \, $
which differ by a reciprocal lattice vector represent the same
physical state for a given conduction band. In this approach
the dispersion relation $\, \epsilon ({\bf p}) \, $ should be
considered as a three-periodic function in $\, \mathbb{R}^{3} \, $
with the periods $\, {\bf a}_{1} \, $, $\, {\bf a}_{2} \, $,
and $\, {\bf a}_{3} \, $. The Fermi surface represents in this case
a three-periodic surface in $\, \mathbb{R}^{3} \, $ which can
have in general a rather complicated form 
(see e.g. Fig. \ref{FermiSurf}).

 In the quasiclassical approximation the evolution of the electron
states in metal in the presence of electric and magnetic fields
can be described by the adiabatic system
$${\dot {\bf p}} \,\,\,\, = \,\,\,\, {e \over c} \,\,
\left[ {\bf v}_{\rm gr} ({\bf p}) \, \times \, {\bf B} \right]
\,\,\, + \,\,\, e \, {\bf E} $$

 In the limit of the strong magnetic fields the behavior of
conductivity is actually defined by the geometry of the dynamical
system
\begin{equation}
\label{MFSyst}
{\dot {\bf p}} \,\,\,\, = \,\,\,\, {e \over c} \,\,
\left[ {\bf v}_{\rm gr} ({\bf p}) \, \times \, {\bf B} \right]
\,\,\,\, = \,\,\,\, {e \over c} \,\, \left[ \nabla \epsilon ({\bf p})
\, \times \, {\bf B} \right]
\end{equation}
which will play the basic role in our considerations here.

\begin{figure}[t]
\begin{center}
\includegraphics[width=0.9\linewidth]{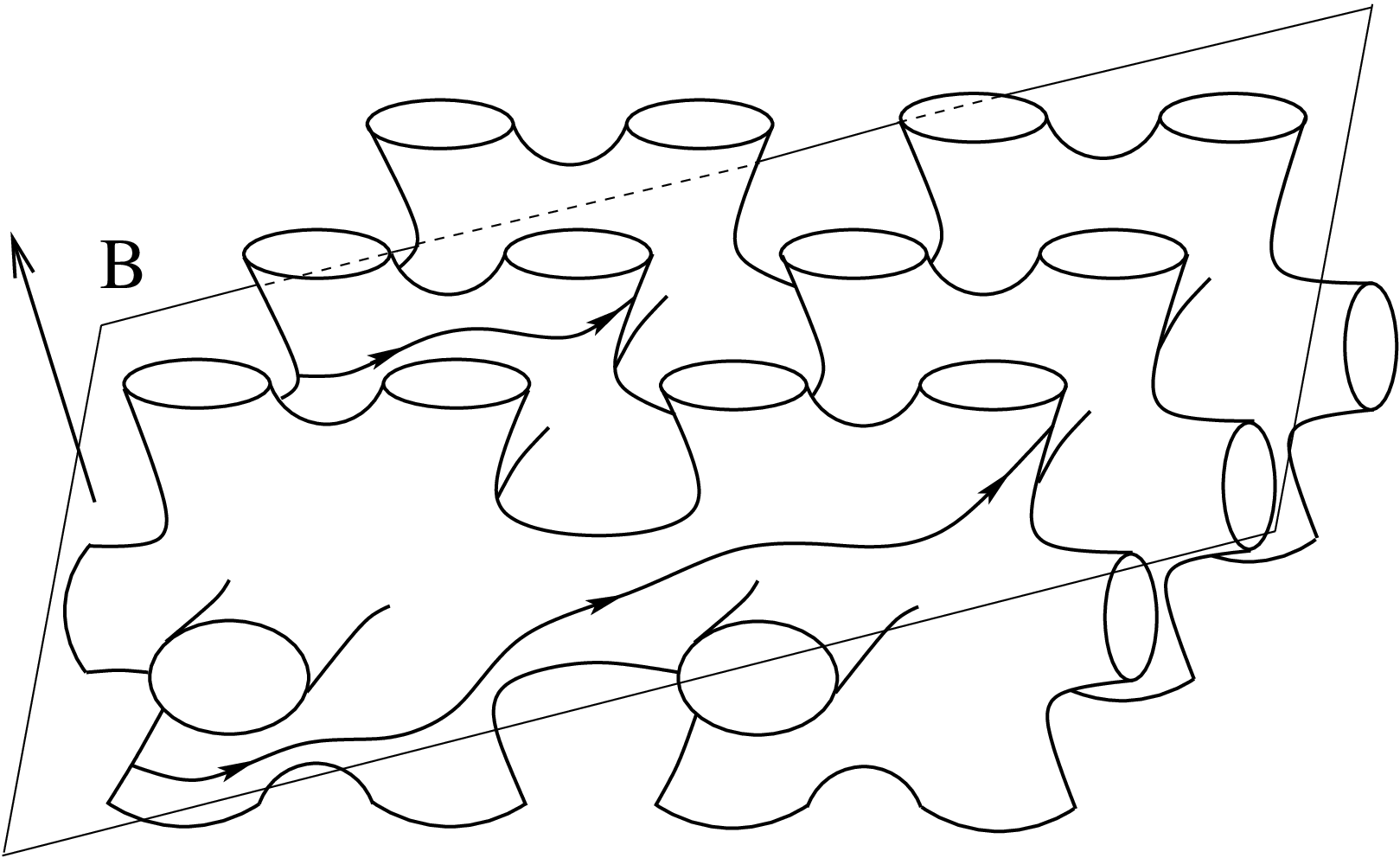}
\end{center}
\caption{An intersection of a complicated Fermi surface by a plane
of general direction in the ${\bf p}$ - space.}
\label{InterFermiSurf}
\end{figure}

 System (\ref{MFSyst}) is integrable from analytical point of view 
and its trajectories are given by  the intersections of the constant
energy levels $\, \epsilon ({\bf p}) \, = \, {\rm const} \, $
with the planes, orthogonal to $\, {\bf B} $. At the same time,
the global geometry of the trajectories of (\ref{MFSyst}) can still
be rather nontrivial in the ${\bf p}$ - space, which can be seen 
if we consider an intersection of a complicated Fermi surface by an
arbitrary plane (Fig. \ref{InterFermiSurf}).

\begin{figure}[t]
\begin{center}
\vspace{0.5cm}
\includegraphics[width=0.9\linewidth]{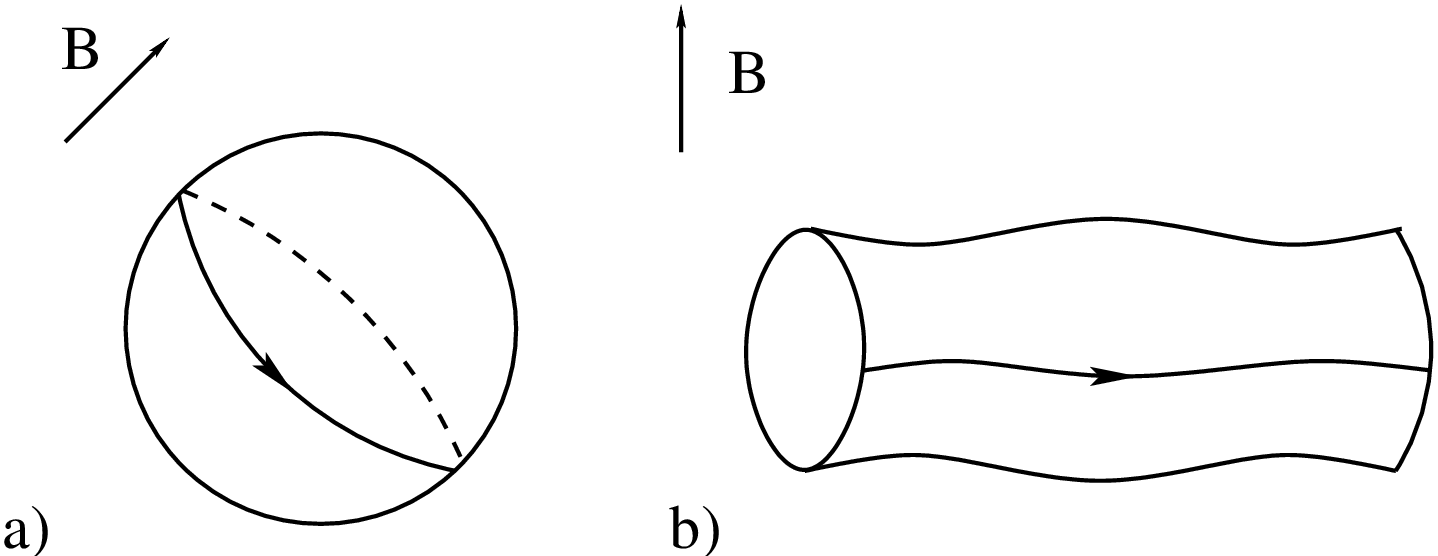}
\end{center}
\caption{Closed (a) and open periodic (b) electron trajectories 
on the Fermi surfaces of different forms.}
\label{ClosedAndPer}
\end{figure}

 An important role of the geometry of open trajectories of system 
(\ref{MFSyst}) for the behavior of conductivity in strong magnetic
fields was first revealed by the school of I.M. Lifshitz
(I.M. Lifshitz, M.Ya. Azbel, M.I. Kaganov, V.G. Peschasky)
in 1950's (see \cite{lifazkag,lifpes1,lifpes2,lifkag1,lifkag2,
lifkag3,etm,KaganovPeschansky}). Thus, in the paper \cite{lifazkag}
the crucial difference between the contribution of the closed and
open periodic trajectories (Fig. \ref{ClosedAndPer})
of (\ref{MFSyst}) to the conductivity tensor in the limit 
$\, \omega_{B} \tau \, \rightarrow \, \infty \, $
was first described.

In papers \cite{lifpes1,lifpes2} more general
types of open trajectories, which are not periodic in 
${\bf p}$ - space and are locally stable with respect to small
rotations of $\, {\bf B} \, $ were considered. Both the open 
periodic trajectories of (\ref{MFSyst}) and more general trajectories,
considered in \cite{lifpes1,lifpes2}, have a mean direction in the
${\bf p}$ - space which results in strong anisotropy of their
contribution to the conductivity in the plane orthogonal to 
$\, {\bf B} \, $. Thus, if we chose the $z$ - axis along the 
direction of $\, {\bf B} \, $ and the $x$ - axis along the mean 
direction of the open trajectories in ${\bf p}$ - space we can 
write for the limiting values of the conductivity tensor
$\, \sigma^{kl} (B) \, $:
\begin{equation}
\label{AnisTen}
\sigma^{kl}_{\infty} \,\,\,\,  =  \,\,\,\,
{n e^{2} \tau \over m^{*}} \,\, \left(
\begin{array}{ccc}
 0  &  0  &  0  \cr
 0  &  *  &  *  \cr
 0  &  *  &  *
\end{array}  \right)
\end{equation}

 In formula (\ref{AnisTen}) the value $\, n \, $ denotes the mean
concentration of the conductivity electrons in metal and 
$\, m^{*} \, $ has a meaning of the effective electron mass in the
crystal. The value $\, \tau \, $ has a meaning of the mean free
electron motion time, and $\, * \, $ denote just some dimensionless
constants of order of unity. Let us note that the projection of a
quasiclassical electron trajectory in ${\bf x}$ - space on the
plane orthogonal to $\, {\bf B} \, $ coincides with the corresponding
trajectory in ${\bf p}$ - space, rotated by $90^{\circ}$, which
explains the form of the tensor (\ref{AnisTen}). It's not
difficult to see that the measurement of the conductivity in the
plane orthogonal to $\, {\bf B} \, $ gives a possibility to find
the mean direction of the open trajectories in ${\bf p}$ - space,
which coincides with the direction of the lowest conductivity in the
limit $\, \omega_{B} \tau \, \rightarrow \, \infty \, $.

 For comparison, the contribution of the closed trajectories of 
system (\ref{MFSyst}) to the conductivity tensor is almost isotropic
in the plane orthogonal to $\, {\bf B} \, $
in the limit $\, \omega_{B} \tau \, \rightarrow \, \infty \, $
and we can write for its limiting values
$$\sigma^{kl}_{\infty} \,\,\,\,  =  \,\,\,\,
{n e^{2} \tau \over m^{*}} \,\, \left(
\begin{array}{ccc}
 0  &  0  &  0  \cr
 0  &  0  &  0  \cr
 0  &  0  &  *
\end{array}  \right) \quad ,  $$
where $\,\, {\hat z} \, = \, {\bf B}/B $.

\begin{figure}[t]
\begin{center}
\vspace{5mm}
\includegraphics[width=0.9\linewidth]{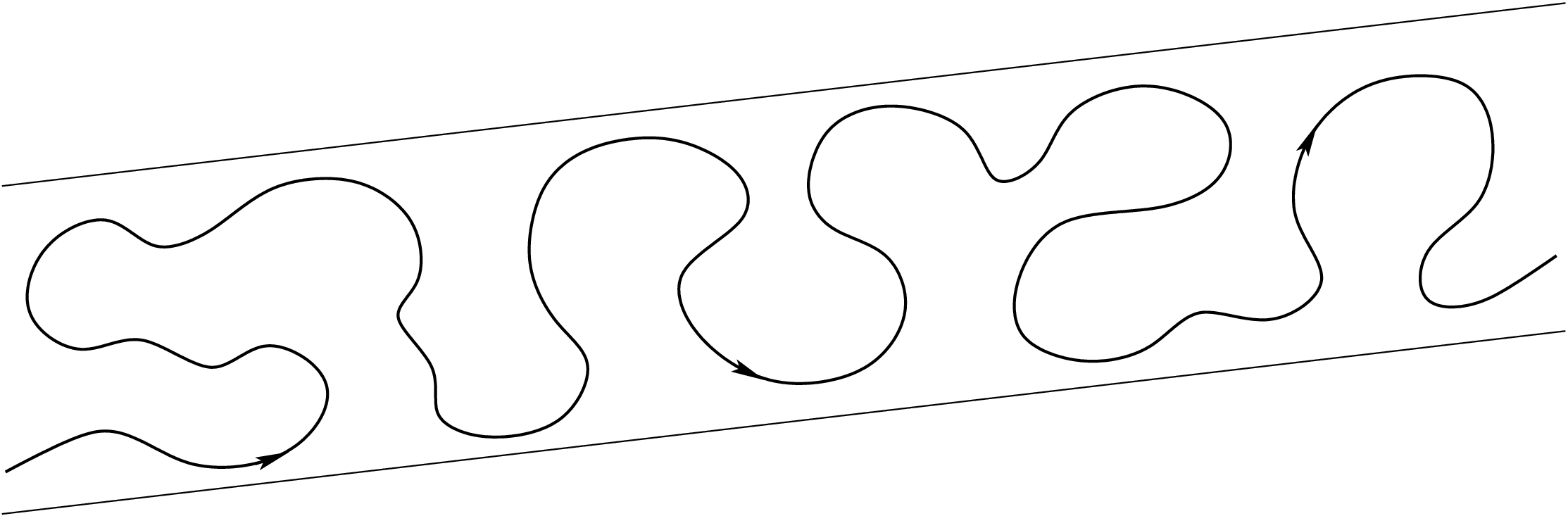}
\end{center}
\caption{The form of a stable open trajectory of system 
(\ref{MFSyst}) in the plane orthogonal to $\, {\bf B} \, $.}
\label{StableTr}
\end{figure}

 The general problem of classification of different trajectories
of system (\ref{MFSyst}) with arbitrary periodic dispersion relation
$\, \epsilon ({\bf p}) \, $ was set by S.P. Novikov
(\cite{MultValAnMorseTheory}) and was intensively studied in his
topological school (S.P. Novikov, A.V. Zorich, S.P. Tsarev, 
I.A. Dynnikov). Let us say here, that this problem turned out 
to be highly non-trivial in its general form and required a set of 
rather deep topological results for its complete investigation
(see \cite{zorich1,dynn1992,Tsarev,dynn1,dynn2}).
The most important achievements in the study of this problem
were made in the papers \cite{zorich1,dynn1} where deep 
topological theorems about the behavior of trajectories of system 
(\ref{MFSyst}) were proved. In particular, the results obtained
in \cite{zorich1} and \cite{dynn1} give a basis for description
of the stable (regular) non-closed trajectories of system
(\ref{MFSyst}) with arbitrary $\, \epsilon ({\bf p}) \, $
which will be also considered in the present paper. Let us formulate 
here the properties of the stable
(with respect to small rotations of $\, {\bf B} \, $ or small
variations of the Fermi energy $\, \epsilon_{F}$) open trajectories
of system (\ref{MFSyst}), which play, from our point of view,
the most important role in the magneto-transport phenomena 
in normal metals:

\vspace{2mm}

1) Every stable open trajectory of system (\ref{MFSyst}) lies
in a straight strip of a finite width in the plane orthogonal to
$\, {\bf B} \, $ and passes through it from $\, - \infty \, $
to $\, + \infty \, $ (Fig. \ref{StableTr});

\vspace{2mm}

2) All the stable open trajectories at a given direction of
$\, {\bf B} \, $ have the same mean direction in ${\bf p}$ - space,
which is given by the intersection of the plane, orthogonal to
$\, {\bf B} \, $, and some locally stable integral plane in the
${\bf p}$ - space.

\vspace{2mm}

 The properties formulated above were used in \cite{PismaZhETF}
for the introduction of important topological characteristics
of electron spectra in metals observable in the transport phenomena
in strong magnetic fields. These characteristics were called in
\cite{PismaZhETF} the topological quantum numbers observable in the 
conductivity of normal metals and can be described in the following 
way:

\vspace{1mm}

 First, according to property (1) we should observe a strong
anisotropy of conductivity in the plane orthogonal to 
$\, {\bf B} \, $ in the limit
$\, \omega_{B} \tau \, \rightarrow \, \infty \, $
in the case of presence of stable open trajectories on the
Fermi surface. The limiting values of the conductivity tensor
$\, \sigma^{kl} (B) \, $ are given by the formula (\ref{AnisTen})
and we can define the mean direction of the open trajectories
of (\ref{MFSyst}) as the direction of the lowest conductivity
in the plane orthogonal to $\, {\bf B} \, $ for
$\, \omega_{B} \tau \gg 1 $. Due to the stability properties of the
open trajectories we can define also these directions for close
directions of $\, {\bf B} \, $ and define an integral plane 
$\, \Gamma \, $, which is swept by the directions of the lowest
conductivity in a given ``Stability Zone'' in the space of
directions of $\, {\bf B} \, $.

\vspace{1mm}

 Let us mention here that integral character of the plane 
$\, \Gamma \, $ in the ${\bf p}$ - space means that it is generated 
by some two reciprocal lattice vectors $\, {\bf q}_{1}$,
$\, {\bf q}_{2} $:
$$\begin{array}{c}
\Gamma \,\,\, = \,\,\, \{ \lambda \, {\bf q}_{1} \,\, + \,\,
\mu \, {\bf q}_{2} \, , \quad \lambda, \mu \, \in \, \mathbb{R} \}
\,\,\, ,   \\  \\
{\bf q}_{1} \,\, = \,\, n_{1} {\bf a}_{1} \, + \, n_{2} {\bf a}_{2}
\, + \, n_{3} {\bf a}_{3} \,\,\, ,  \\  \\
{\bf q}_{2} \,\, = \,\,	m_{1} {\bf a}_{1} \, + \, m_{2} {\bf a}_{2}
\, + \, m_{3} {\bf a}_{3} \,\,\, ,   \\  \\
(n_{1}, n_{2}, n_{3}, m_{1}, m_{2}, m_{3} \, \in \, \mathbb{Z})  
\end{array}  $$

 In the ${\bf x}$ - space the plane $\, \Gamma \, $ can be given
by an indivisible triple of integers $\, (M_{1}, M_{2}, M_{3}) \, $
from the equation
$$M_{1} \, \left( {\bf x} , {\bf l}_{1} \right) \,\, + \,\, 
M_{2}	\, \left( {\bf x} , {\bf l}_{2}	\right)	\,\, + \,\,  
M_{3}	\, \left( {\bf x} , {\bf l}_{3}	\right)	\,\,\, = \,\,\, 0 $$
where $\, ({\bf l}_{1}, {\bf l}_{2}, {\bf l}_{3}) \, $ represent
the basis of the direct lattice. The numbers
$\, (M_{1}, M_{2}, M_{3}) \, $ were called in \cite{PismaZhETF}
the topological quantum numbers observable in conductivity of
normal metals and represent the homology classes of two-dimensional
``carriers of open trajectories'' in the torus 
$\, \mathbb{T}^{3} \, $. Let us say that the triples
$\, (M_{1}, M_{2}, M_{3}) \, $ can be rather nontrivial for
complicated Fermi surfaces and represent (together with the
geometry of the ``Stability Zones'') an important characteristic
of the electron spectrum in a metal.

 Another important property of the stable open trajectories of
system (\ref{MFSyst}) is that they never appear together with more
complicated (unstable) chaotic open trajectories (chaotic
trajectories of Tsarev or Dynnikov type) at the same direction of
$\, {\bf B} \, $ (\cite{dynn3}). As a result, the contribution
of the trajectories shown at Fig. \ref{StableTr} to the conductivity
represents the only nontrivial part of the tensor 
$\, \sigma^{kl} ({\bf B}) \, $ in the ``Stability Zone'' and is 
easily observable in experiments. Let us say here also, that the 
consideration of ``chaotic'' trajectories of system (\ref{MFSyst})
will not be a subject of the present paper.

 The most detailed mathematical survey on the geometry of
trajectories of system (\ref{MFSyst}) is represented in the paper
\cite{dynn3}. The detailed description of the physical phenomena
based on topological investigations of the system (\ref{MFSyst})
can be found in the papers \cite{UFN,BullBrazMathSoc,JournStatPhys}.
Let us give also a reference to the paper \cite{DeLeoPhysB}
where a convenient mathematical method of numerical investigation
of the structure of Stability Zones, suggested by I.A. Dynnikov,
was used by R. De Leo for investigation of Stability Zones for
a set of analytical dispersion relations, used as approximations
to dispersion relations in real crystals. We can give here also 
a reference to the papers 
\cite{dynn2,zorich2,ZhETF1997,DeLeo1,DeLeo2,DeLeo3,DeLeoDynnikov1,
DeLeoDynnikov2,Skripchenko1,Skripchenko2,DynnSkrip} 
devoted to investigation of different
aspects of chaotic trajectories of system (\ref{MFSyst}), which
can arise on rather complicated Fermi surfaces.

\begin{figure}[t]
\begin{center}
\vspace{5mm}
\includegraphics[width=0.9\linewidth]{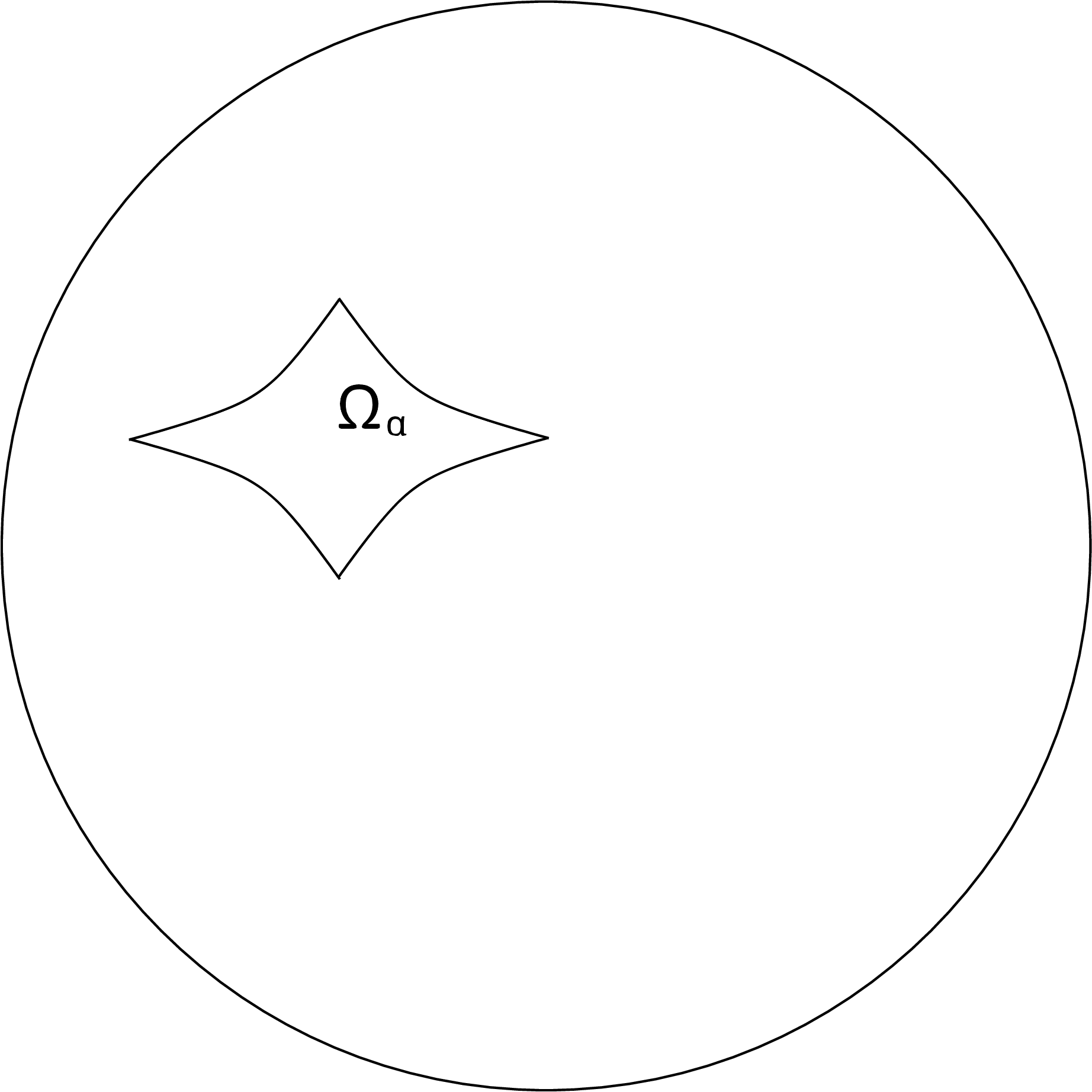}
\end{center}
\caption{An exact mathematical Stability Zone $\, \Omega_{\alpha} \, $
on the unit sphere $\, \mathbb{S}^{2} \, $, corresponding to a fixed 
triple $\, (M^{\alpha}_{1}, M^{\alpha}_{2}, M^{\alpha}_{3}) \, $.}
\label{ExactMathZone}
\end{figure}

 The present paper will be devoted to the methods of experimental
determination of the boundaries of exact mathematical
``Stability Zones'', which represents in fact a nontrivial problem
from experimental point of view. Let us say at once that we define
the exact mathematical Stability Zone $\, \Omega_{\alpha} \, $
as a region on the angle diagram (unit sphere 
$\, \mathbb{S}^{2} \, $) corresponding to the presence of the stable 
open trajectories with the same topological quantum numbers
$\, (M^{\alpha}_{1}, M^{\alpha}_{2}, M^{\alpha}_{3}) \, $
on the Fermi surface. According to this definition, the open
trajectories of system (\ref{MFSyst}) exist for any
$\, {\bf B}/B \, \in \, \Omega_{\alpha} \, $,
are stable with respect to small rotations of $\, {\bf B} \, $
and define the same integral plane $\, \Gamma_{\alpha} \, $
in the ${\bf p}$ - space. The Stability Zone $\, \Omega_{\alpha} \, $,
defined in this way, represents a finite region on the unit sphere
$\, \mathbb{S}^{2} \, $ with a piecewise smooth boundary
(Fig. \ref{ExactMathZone}).

\begin{figure}[t]
\begin{center}
\vspace{5mm}
\includegraphics[width=0.9\linewidth]{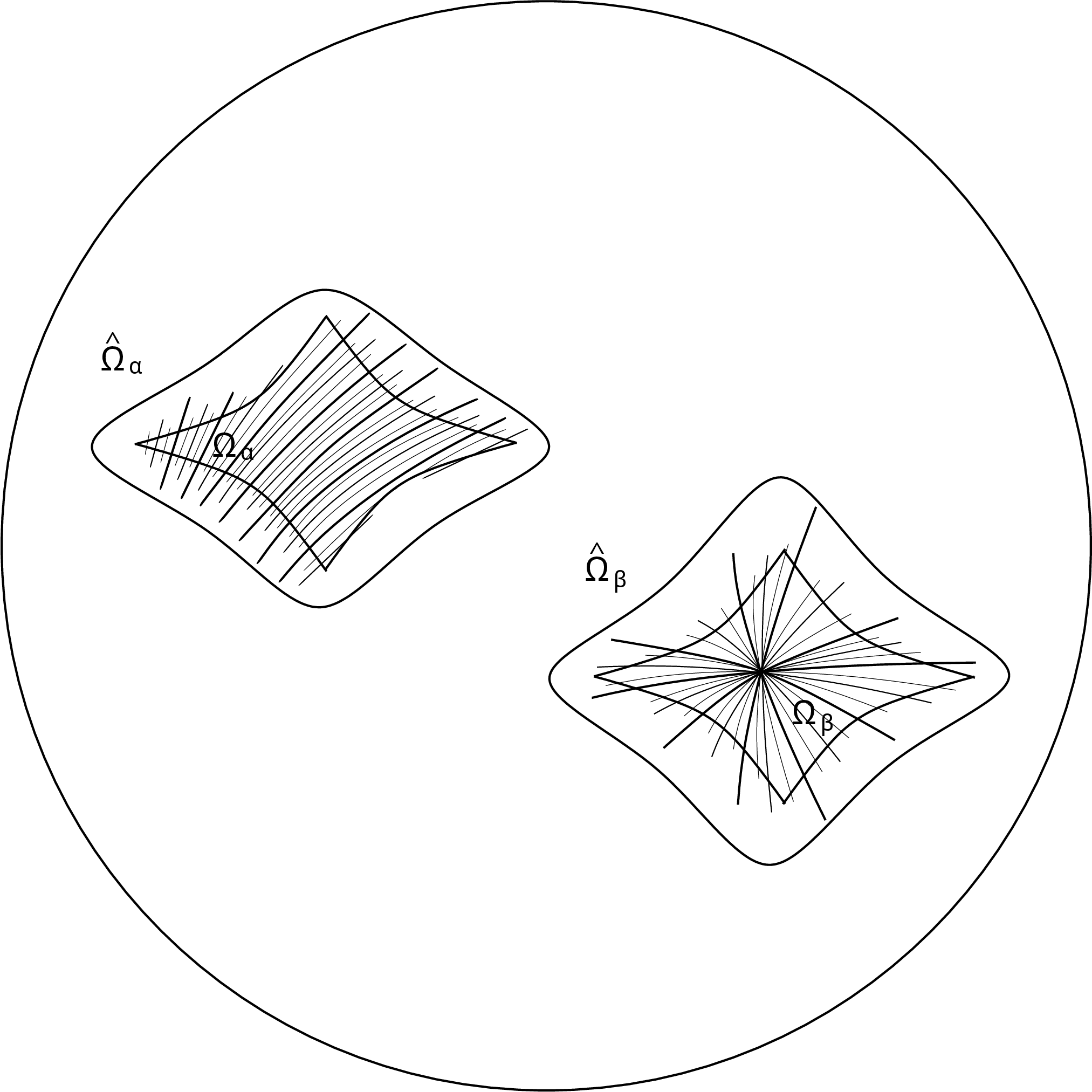}
\end{center}
\caption{The ``experimentally observable Stability Zones''
with nets of special directions of $\, {\bf B} \, $ on the
angle diagram.}
\label{ExpZone}
\end{figure}

 Let us say, however, that the full set of directions of
$\, {\bf B} \, $, corresponding to the presence of the open
trajectories on the Fermi surface, has in general more complicated 
structure. Thus, it was first pointed out in \cite{GurzhyKop} 
that the boundaries of the regions on the angle diagram, corresponding 
to appearance of the open trajectories, should have in fact a singular 
structure, which is caused by the difference between periodic and 
non-periodic trajectories arising on the Fermi surface. Using general 
topological description of the stable open trajectories of system 
(\ref{MFSyst}) it can be shown that a general Stability Zone 
$\, \Omega_{\alpha} \, $ on $\, \mathbb{S}^{2} \, $ has necessarily 
an everywhere dense ``net'' of directions 
$\, {\bf B} \, \in \, \Omega_{\alpha} \, $,
where the stable open trajectories of system (\ref{MFSyst})
are actually periodic (\cite{JETP2017}). Moreover, this net 
should be in fact extended outside the Zone 
$\, \Omega_{\alpha} \, $, since the periodic open trajectories 
still exist on its segments near the boundary of 
$\, \Omega_{\alpha} \, $. Let us note here, that according to our 
definition we don't include the  corresponding segments into the 
Zone $\, \Omega_{\alpha} \, $ since the corresponding trajectories
are not stable anymore with respect to small rotations of
$\, {\bf B} \, $. Besides that, the closed electron trajectories
near the boundary of $\, \Omega_{\alpha} \, $ have in fact very
specific form which makes them hardly distinguishable from the 
open trajectories from experimental point of view. As a result,
the ``experimentally observable Stability Zone'' 
$\, {\hat \Omega}_{\alpha} \, $ is actually different from the 
exact mathematical Stability Zone (Fig. \ref{ExpZone}).

 Let us say also that due to the difference between the periodic
and non-periodic trajectories the analytic dependence of the
values $\, \sigma^{kl} ({\bf B}) \, $ both on the value and the
direction of $\, {\bf B} \, $ is actually rather complicated both 
in the Zone $\, \Omega_{\alpha} \, $ and 
$\, {\hat \Omega}_{\alpha} \, $ (see \cite{JETP2017}).
As a result, the exact boundary of a mathematical Stability
Zone $\, \Omega_{\alpha} \, $ is in fact unobservable in direct
measurements of the values $\, \sigma^{kl} ({\bf B}) \, $ even
for rather big values of $\, B \, $.

 On the other hand, the exact form of the mathematical 
Stability Zones $\, \Omega_{\alpha} \, $ represents an important 
characteristic of the dispersion relation 
$\, \epsilon({\bf p}) \, $ and can play rather important role in 
the reconstruction of the form of the Fermi surface from 
experimental data. In the present paper we will show that the 
oscillation phenomena in normal metals give in fact a convenient
way to determine the boundary of the Zones $\, \Omega_{\alpha} \, $,
which is based on the general topological structure of the
``carriers of  open trajectories'' on the Fermi surface. 
In the next chapter we will give a description of the topological
structure of a (complicated) Fermi surface in the presence of
the stable open trajectories of system (\ref{MFSyst}) and discuss
special features of the oscillation	phenomena on the surfaces
of this kind.

\vspace{5mm}

\section{Special topological representation of a Fermi surface 
containing stable open trajectories and the oscillation phenomena 
in normal metals.}
\setcounter{equation}{0}

\begin{figure}[t]
\begin{center}
\vspace{5mm}
\includegraphics[width=0.9\linewidth]{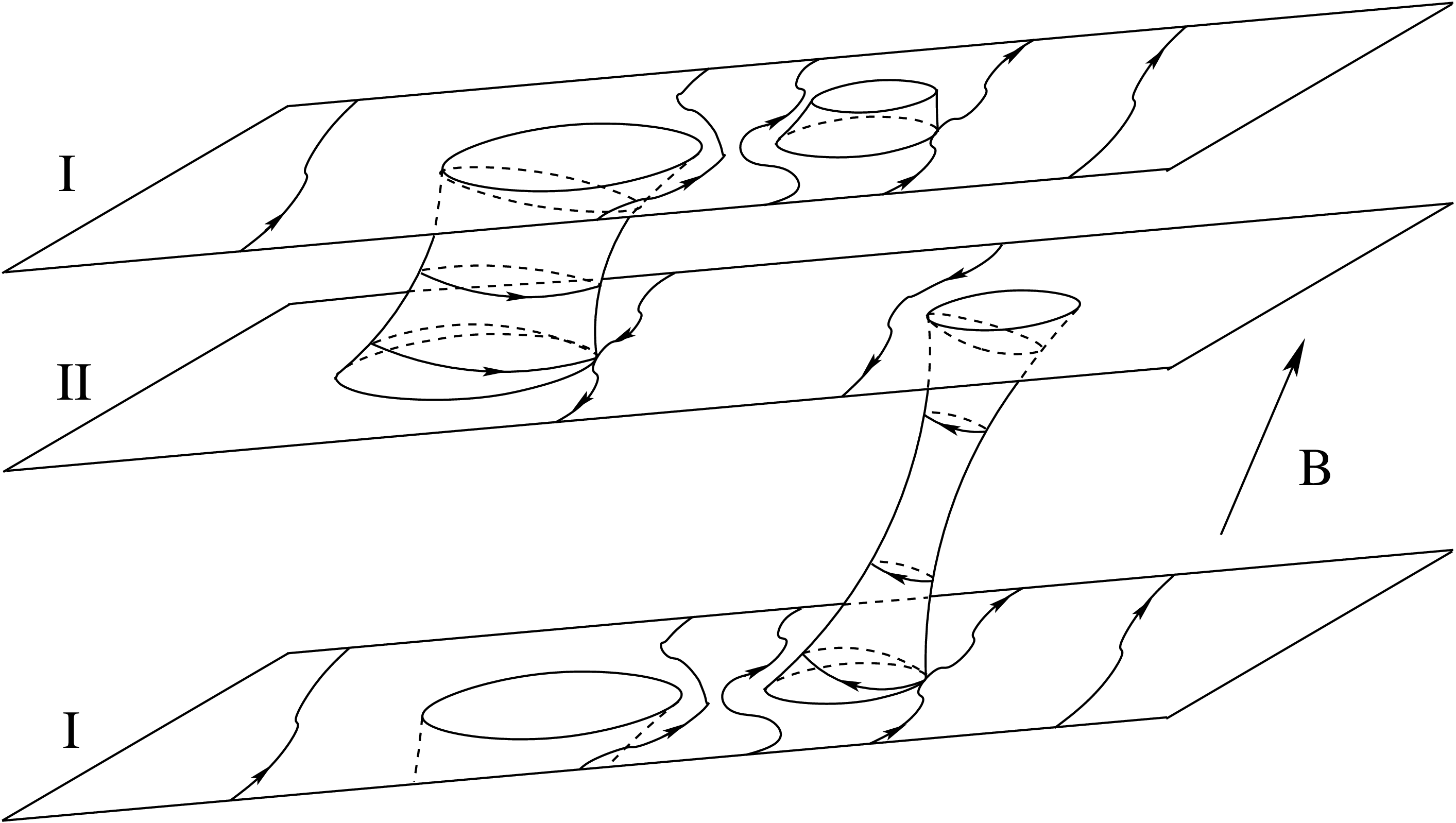}
\end{center}
\caption{A model Fermi surface of genus 3 in ${\bf p}$ - space
with stable open trajectories of system (\ref{MFSyst}).}
\label{Gen3FermiSurf}
\end{figure}

 To describe the special topological representation of a complicated 
Fermi surface in presence of the stable open trajectories of
system (\ref{MFSyst}) let us introduce first a model Fermi surface,
having the following form:

 Consider a periodic set of parallel integral planes in
${\bf p}$ - space, connected by cylinders of finite heights
(Fig. \ref{Gen3FermiSurf}).

 Let us divide all the planes into two different classes
(I and II) - the odd-numbered planes and the even-numbered planes.
Let us assume now that the planes of each class can be obtained
from each other by a shift on some reciprocal lattice vector and
represent the same object after the factorization over the 
reciprocal lattice.

 In the same way, we assume that the cylinders are also divided
into two classes which represent just two non-equivalent objects
after the factorization.

 We can say then that the periodic surface described above gives
an example of a Fermi surface, which is represented as a pair of 
two parallel two-dimensional tori 
$\, \mathbb{T}^{2} \, \subset \, \mathbb{T}^{3} \, $
embedded in $\, \mathbb{T}^{3} \, $ and two cylinders of finite
heights, connecting the tori $\, \mathbb{T}^{2} \, $.

 Let us assume now that the axes of the cylinders are almost
parallel in ${\bf p}$ - space and consider a magnetic field
having a direction close to the direction of the axes of the
cylinders. It is not difficult to see that if the direction 
of $\, {\bf B} \, $ is almost parallel to the cylinders 
then the cylinders consist mostly of closed trajectories
of system (\ref{MFSyst}), which cut our Fermi surface into
separate parallel (deformed) planes. It can be also noted that the 
closed trajectories arising on the cylinders of different types
also belong to different (the electron-type or the hole-type)
types. The open trajectories of system (\ref{MFSyst}) are given
by the intersections of the planes, orthogonal to $\, {\bf B} \, $,
with the periodically deformed integral planes in ${\bf p}$ - space
and have the regular form shown at Fig. \ref{StableTr}.
The parts of the Fermi surface, consisting of closed trajectories,
represent cylinders, restricted by singular trajectories, with heights, 
depending on the direction of $\, {\bf B} \, $. Thus, we have 
a Stability Zone around our initial direction of $\, {\bf B} \, $
with topological quantum numbers, defined by the homology class
of the integral planes introduced above.

\begin{figure}[t]
\begin{center}
\vspace{5mm}
\includegraphics[width=0.9\linewidth]{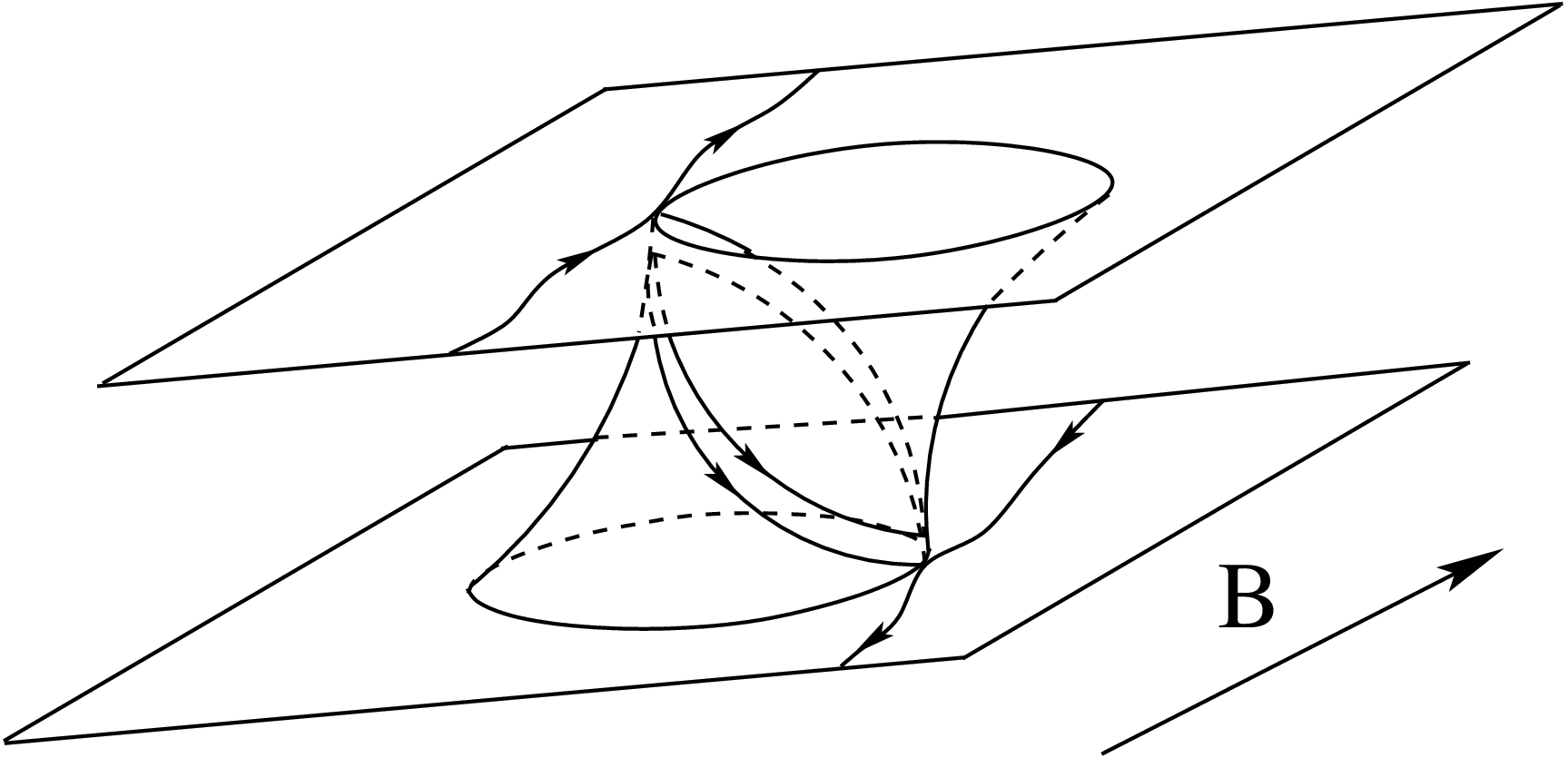}
\end{center}
\caption{A very short cylinder of closed trajectories of one
type for a direction of $\, {\bf B} \, $, close to the boundary
of a Stability Zone.}
\label{Cylind}
\end{figure}

 The boundary of the Stability Zone is defined by the condition
that the height of the cylinders of one type becomes zero, such
that the corresponding closed trajectories disappear on the Fermi
surface (Fig. \ref{Cylind}). As a result, the remaining closed
trajectories can not cut the Fermi surface anymore into separate
planes and we do not have stable open trajectories after the 
crossing the boundary of a Stability Zone. Thus, a
Stability Zone represents in general a region with a piecewise
smooth boundary in the space of directions of $\, {\bf B} \, $.

 Let us note, that we don't claim here that the open trajectories
of system (\ref{MFSyst}) completely disappear outside the
Stability Zone $\, \Omega \, $. Indeed, we can see that the
remaining cylinders of closed trajectories now cut the 
Fermi surface into connected pairs of integral planes and the
trajectories of system (\ref{MFSyst}) still admit an effective
description near the boundary of the Zone $\, \Omega \, $.
It is not difficult to see, that the trajectories can now 
``jump'' between two planes which gives a reconstruction of
the open trajectories after the crossing of the boundary
of the Zone $\, \Omega \, $. It can be seen also that all the 
open trajectories will transform into long closed trajectories
if the intersection of the plane, orthogonal to $\, {\bf B} \, $,
with the integral planes has an irrational direction in the
${\bf p}$ - space. At the same time, if the intersection
of the plane, orthogonal to $\, {\bf B} \, $, and the integral
planes has a rational direction, then we will have both the
long closed trajectories and the open periodic trajectories
near the boundary of the Zone $\, \Omega \, $. The open periodic
trajectories, however, are not stable outside the Zone
$\, \Omega \, $, so we should not include the corresponding
set of directions of $\, {\bf B} \, $ in the mathematical
Stability Zone. At the same time, this set belongs to the
``experimentally observable Stability Zone'', which includes
the mathematical Stability Zone as a subset. The analytical
properties of the conductivity in the experimentally
observable Stability Zone are in fact rather complicated
(see \cite{JETP2017}), which makes the experimental determination
of the boundary of a mathematical Stability Zone a non-trivial
problem.

 The picture described above represents a Fermi surface
of genus 3, embedded in the three-dimensional torus
$\, \mathbb{T}^{3} \, $, and gives an example of complicated
enough Fermi surface from topological point of view. 
In general, the topological representation of real complicated
Fermi surface, carrying stable open trajectories of system
(\ref{MFSyst}), can differ from that described above in the
following details:

\vspace{2mm}

1) The number of non-equivalent cylinders of closed
trajectories can be bigger (or less) than 2;

2) There can be additional cylinders of closed trajectories
on the integral planes, having a point as a base;

\begin{figure}[t]
\begin{center}
\vspace{5mm}
\includegraphics[width=\linewidth]{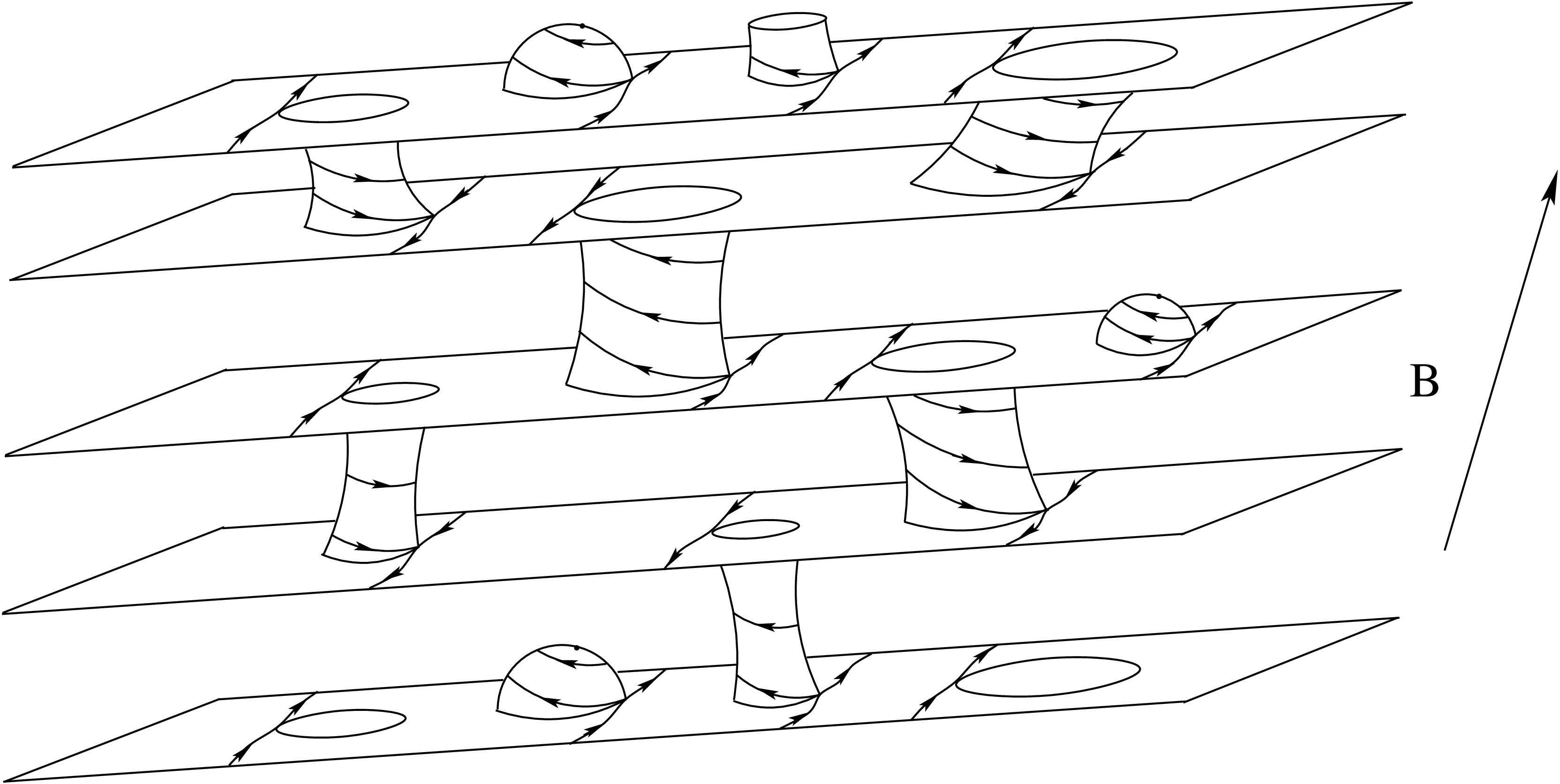}
\end{center}
\caption{Topological representation of a Fermi surface of 
a very high genus carrying stable open trajectories of system 
(\ref{MFSyst}).}
\label{ComplFermiSurf}
\end{figure}

3) The number of non-equivalent parallel integral planes
can be bigger than 2, being any even number 
(Fig. \ref{ComplFermiSurf}).

\vspace{2mm}

 Let us say here that in the most general case we can also assume 
that parts of the Fermi surface consisting of closed trajectories 
and connecting carriers of open trajectories can have a composite 
structure and consist of several cylinders of closed trajectories 
inside some Stabilty Zone $\, \Omega_{\alpha} $. This possibility 
does not change the essence of our further considerations, and is 
extremely unlikely for real Fermi surfaces, so we will not dwell on 
it here. Thus, we can consider the described structure of the Fermi 
surface as the most common near the boundary of Stability Zones.

\vspace{1mm}

\begin{figure}[t]
\begin{center}
\vspace{5mm}
\includegraphics[width=0.9\linewidth]{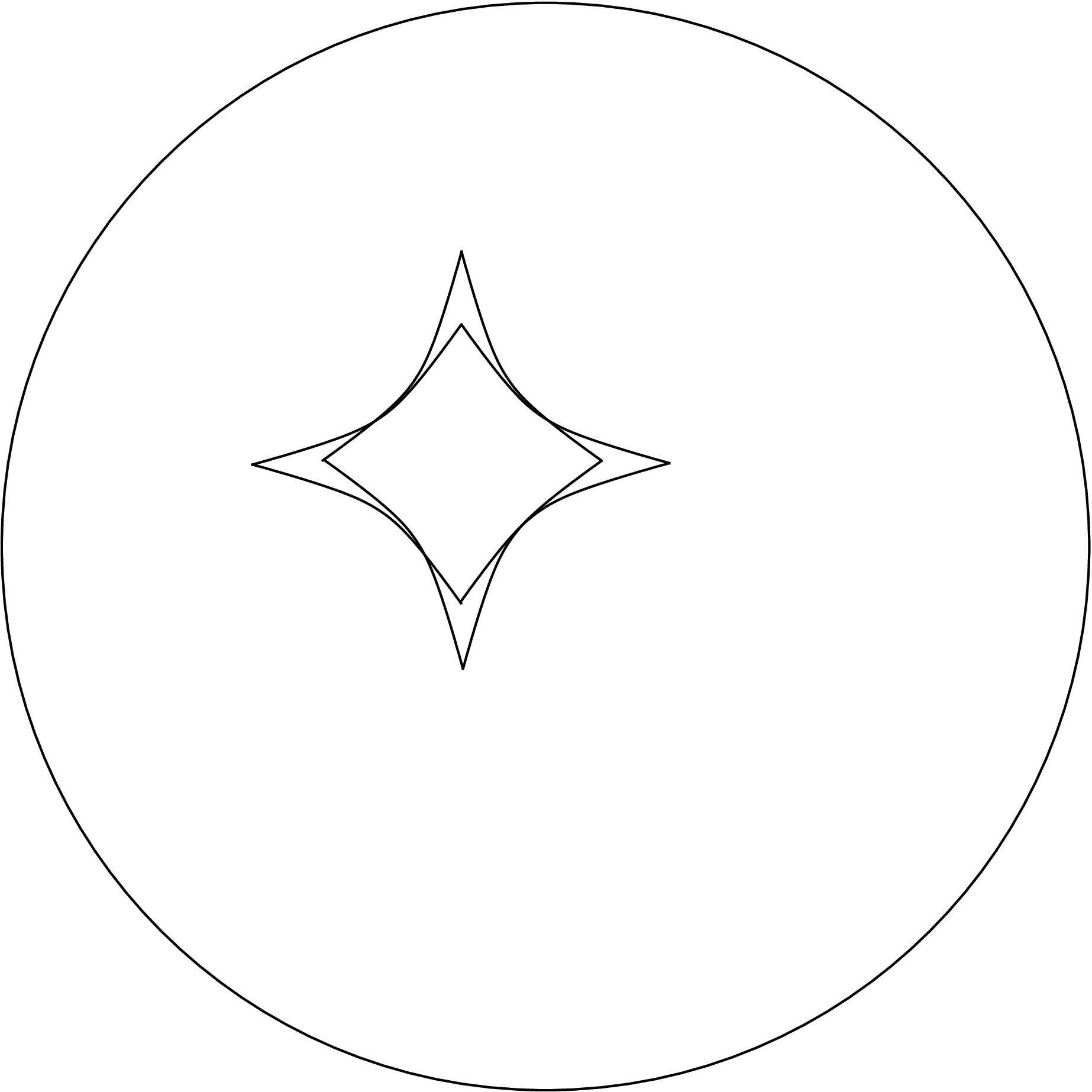}
\end{center}
\caption{A complicated Stability Zone, representing an overlapping
of two Zones with the same topological numbers.}
\label{ComplZone}
\end{figure}

 The general statement formulated above represents a corollary
of rather deep topological theorems, proved in the papers
\cite{zorich1,dynn1,dynn3}. Let us say that the situation (3)
can be observed actually only for surfaces of very high genus,
so, for many real metals it actually does not arise. Let us
note also, that the situation (3) can be considered also 
as an ``overlapping'' of two (or more) Stability Zones with
the same topological quantum numbers. As was pointed out in
\cite{JETP2017}, the analytical properties of conductivity
are the most complicated in this situation, in particular,
we should observe here more than one boundary of a Stability
Zone (Fig. \ref{ComplZone}).

 In any situation the boundary of an exact mathematical Stability
Zone is defined by vanishing of one of the cylinders of closed
trajectories, so, as we will see below, the study of the
oscillation phenomena gives in fact a convenient instrument 
to detect the boundary of a Stability Zone in experiment.

 Let us say also here, that the picture described above 
represents a purely topological representation of a Fermi surface, 
carrying stable open trajectories (or, better to say, a topological 
representation of system (\ref{MFSyst}) on the Fermi surface), and 
can be visually much more complicated due to possible complicated 
geometry of the objects, introduced above.

 Below we will consider the oscillation phenomena
in the picture described above and describe their special features
near the boundary of a Stability Zone. Certainly, we will not give 
here a detailed theoretical exposition of the phenomena we are going 
to consider and give just a reference on their standard description
(see e.g. \cite{etm,Kittel,Abrikosov,Ziman}).

 Let us start with the (classical) cyclotron resonance phenomenon, 
which can be described with the aid of the purely kinetic approach 
for the electron gas in normal metals. As it is well known, the
cyclotron resonance phenomenon is connected with the oscillating
dependence of the surface conductivity on the frequency of the
alternating field in presence of a strong magnetic field
$\, {\bf B} \, $. In the most common setting, the direction of
$\, {\bf B} \, $ is assumed to be parallel to the metal surface
and the alternating electric field can have different directions
in the same plane. The oscillating behavior of the surface 
conductivity (in the situation of anomalous skin effect) is caused 
by the coincidence of the frequency $\, \Omega \, $ of the incident 
wave with the values 
$\, n \, \omega_{B} \, $, $\, n \in \mathbb{N} \, $,
where $\, \omega_{B} \, = \, \omega_{B} (p_{z}) \, $ is 
the cyclotron frequency, defined for every closed trajectory of 
system (\ref{MFSyst}). In general, $\, \omega_{B} (p_{z}) \, $
represents a complicated function of $\, p_{z} \, $ and the 
oscillating behavior of conductivity is determined in fact
by the extremal values of $\, \omega_{B} (p_{z}) \, $,
satisfying the condition $\, d \omega_{B} / d p_{z} \, = \, 0 $.

 In the geometric picture described above the closed trajectories
are combined into cylinders connecting integral planes and we have
the relation $\, \omega_{B} (p_{z}) \, = \, 0 $ on the bases of 
the cylinders. As a result, the positive function 
$\, \omega_{B} (p_{z}) \, $ should have at least one maximum
at every cylinder, i.e. every cylinder of closed trajectories
contains at least one extremal trajectory in the sense 
pointed above. Let us say, that theoretically we can have
several maxima and minima of the function 
$\, \omega_{B} (p_{z}) \, $ on a cylinder of closed trajectories,
however, the situation of several critical points of
$\, \omega_{B} (p_{z}) \, $ on the same cylinder requires 
in fact rather complicated geometry of the dispersion relation. 
For simplicity, we will assume here that all the cylinders
of closed trajectories contain just one extremal (maximal) value 
of $\, \omega_{B} (p_{z}) \, $. As we will see, more complicated
cases do not contain any fundamental differences from the simple 
case under consideration. 

\begin{figure}[t]
\begin{center}
\vspace{5mm}
\includegraphics[width=0.9\linewidth]{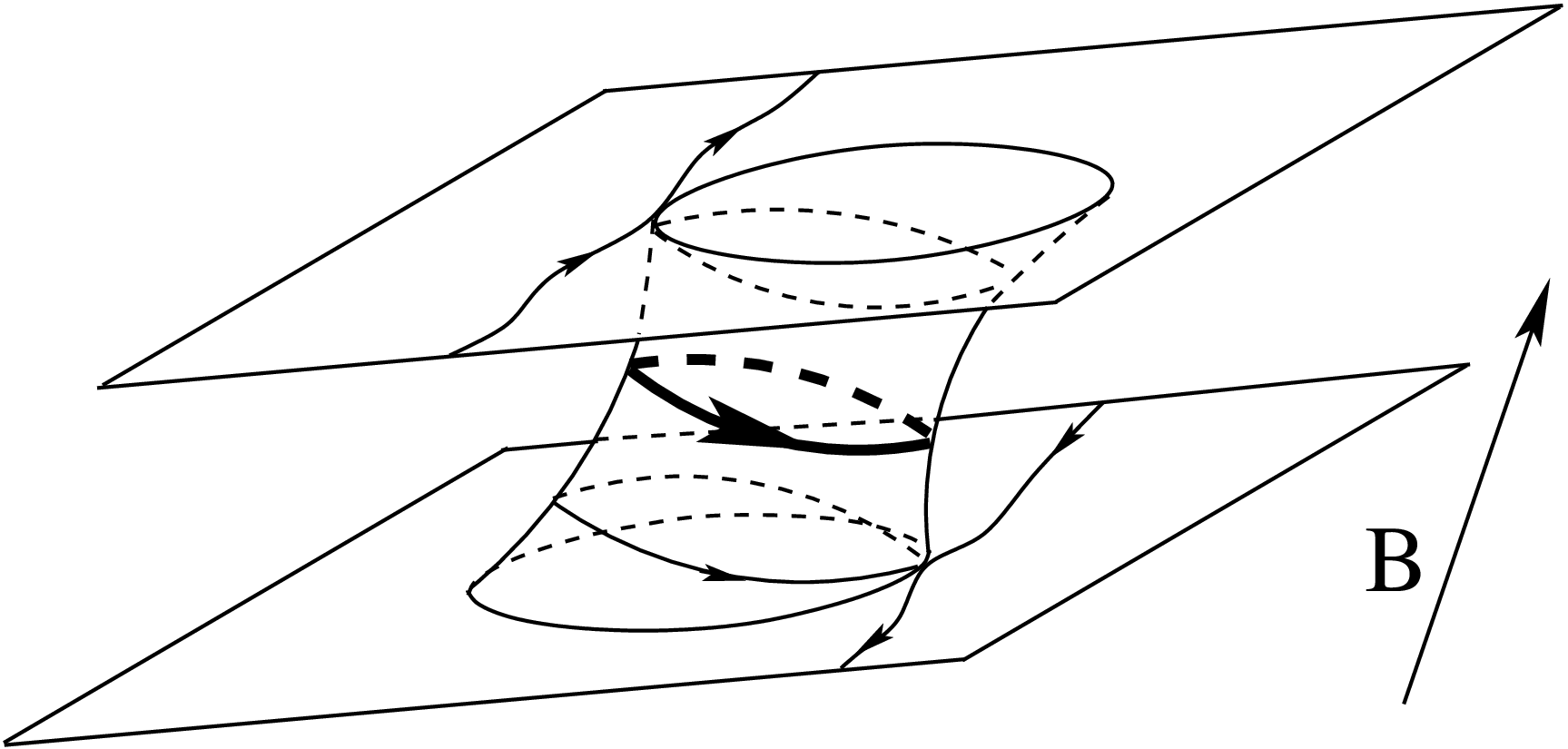}
\end{center}
\caption{The extremal trajectory on a cylinder of closed 
trajectories of system (\ref{MFSyst}).}
\label{CentralSection}
\end{figure}

 Let us make also another remark. For Fermi surfaces of not
very high genus the boundary of a Stability Zone is defined 
by disappearance of just one cylinder of closed trajectories
of system (\ref{MFSyst}), which is invariant under the 
transformation $\, {\bf p} \rightarrow - {\bf p} \, $.
In this case the extremal value of $\, \omega_{B} \, $
corresponds to the central cross-section of the cylinder by 
the plane orthogonal to $\, {\bf B} \, $ 
(Fig. \ref{CentralSection}). For more complicated
Fermi surfaces (of a high genus) we can also have the situation
when two non-equivalent cylinders, which transform into
each other under the transformation 
$\, {\bf p} \rightarrow - {\bf p} \, $,
disappear simultaneously on the boundary of a Stability Zone.
Let us say, however, that this situation requires in fact
a really complicated Fermi surface, so, in most of experiments we 
can actually assume the first case. Let us note also here,
that the first situation has also the additional property
$$\langle v^{z}_{gr} \rangle_{tr} \,\,\, = \,\,\, 0 $$
on the extremal trajectories, which permits in fact not
to impose too strict condition that $\, {\bf B} \, $
is parallel to the metal surface.

\begin{figure}[t]
\begin{center}
\vspace{5mm}
\includegraphics[width=\linewidth]{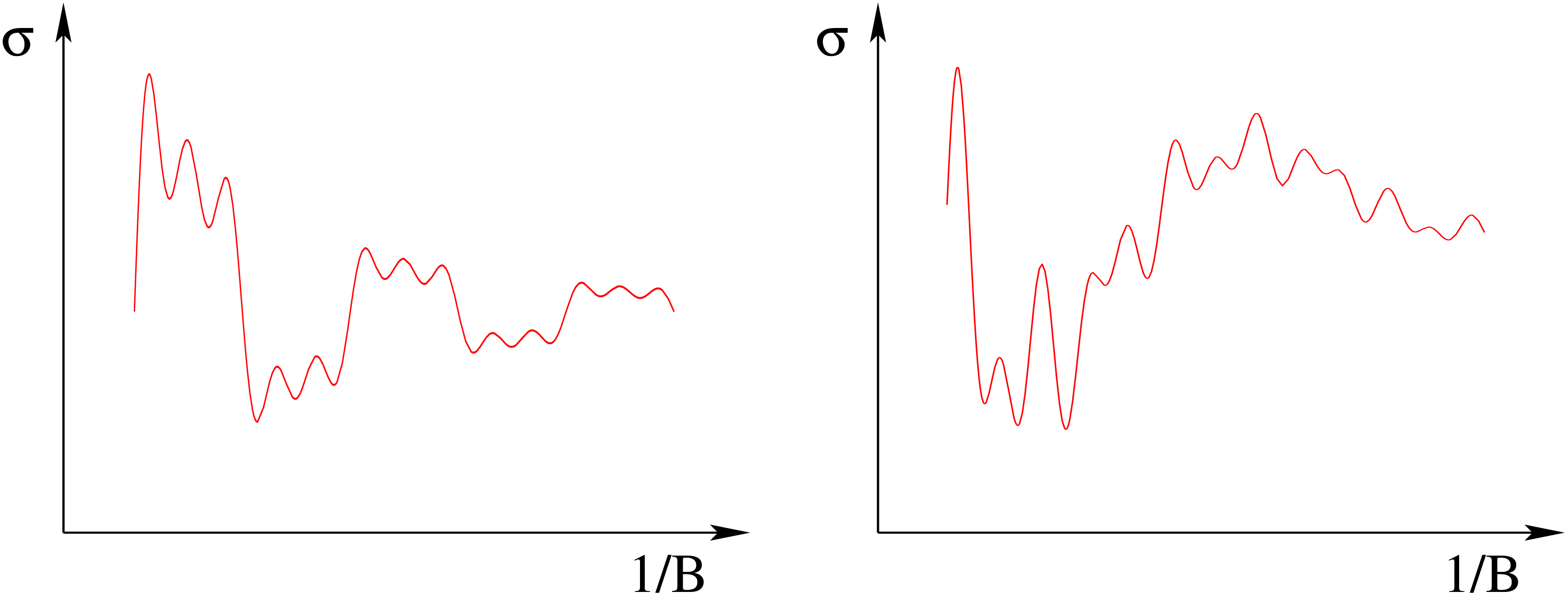}
\end{center}
\caption{The increasing of the period of one of the main
oscillating terms in the surface conductivity 
(real or imaginary part) near the boundary
in a Stability Zone.}
\label{CyclRes1}
\end{figure}

\begin{figure}[t]
\begin{center}
\vspace{5mm}
\includegraphics[width=\linewidth]{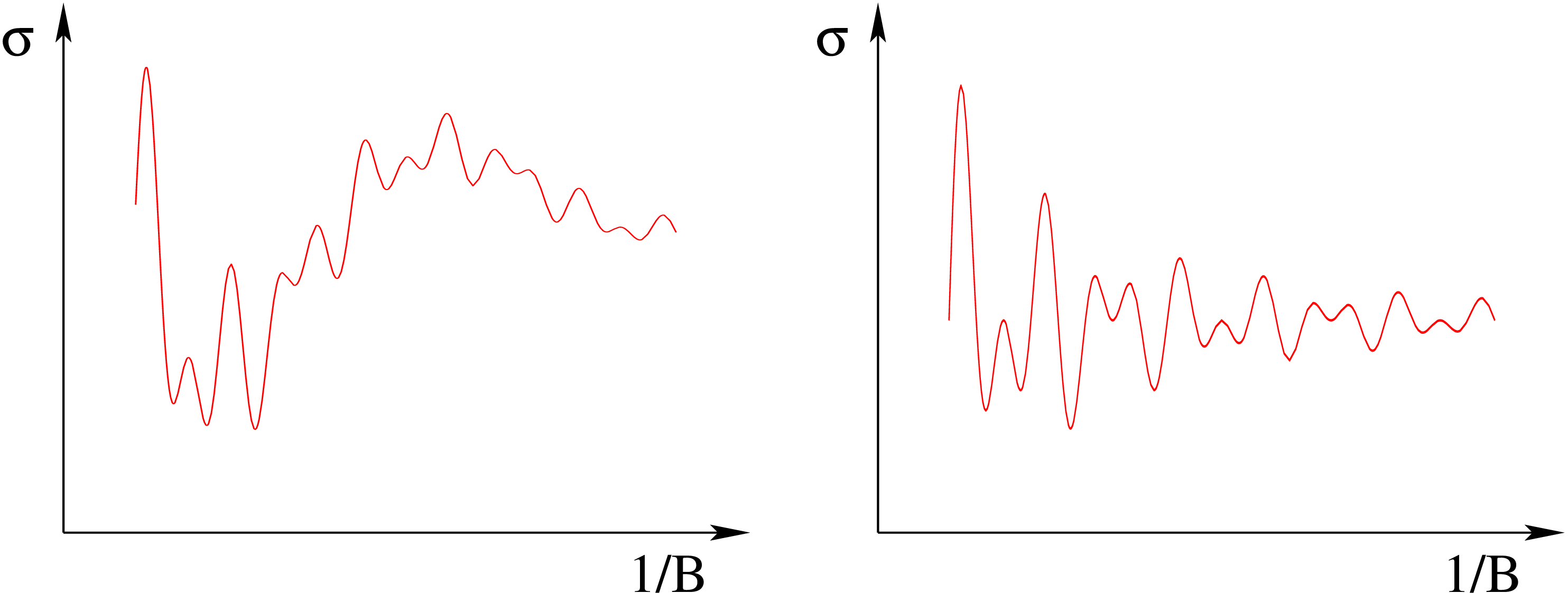}
\end{center}
\caption{The disappearance of one of the main oscillating
contributions in the surface conductivity after 
crossing the boundary of a Stability Zone.}
\label{CyclRes2}
\end{figure}

 Coming back to the cyclotron resonance phenomenon, we can see 
then that for the structure of the Fermi surface described above 
the oscillating conductivity behavior in high frequency electric 
fields should reveal a finite number of the main terms, 
representing the contribution of the extremal trajectories 
on the cylinders of closed trajectories and also of the endpoints 
of the special (degenerate) cylinders at Fig. \ref{ComplFermiSurf}. 
Let us also note here that the last contribution actually does not
arise if the direction of $\, {\bf E} \, $ in the incident
wave is orthogonal to our fixed direction of $\, {\bf B} \, $.
Every main term is characterized by a periodic dependence on the 
frequency $\, \Omega \, $ (or on $\, 1 / B \, $) with its own
period $\, T_{i} \, $. For directions of $\, {\bf B} \, $
lying inside a Stability Zone $\, \Omega_{\alpha} \, $ the periods
$\, T_{i} ({\bf B}/B) \, $ represent smooth functions of the
direction of $\, {\bf B} \, $ and one of the periods 
$\, T_{i_{0}} \, $ becomes infinite at the boundary of the
Stability Zone (Fig. \ref{CyclRes1}). After crossing the boundary 
of a mathematical Stability Zone $\, \Omega_{\alpha} \, $ 
the corresponding contribution disappears completely, while all 
the other main contributions do not notably change 
(Fig. \ref{CyclRes2}).

\begin{figure}[t]
\begin{center}
\vspace{5mm}
\includegraphics[width=\linewidth]{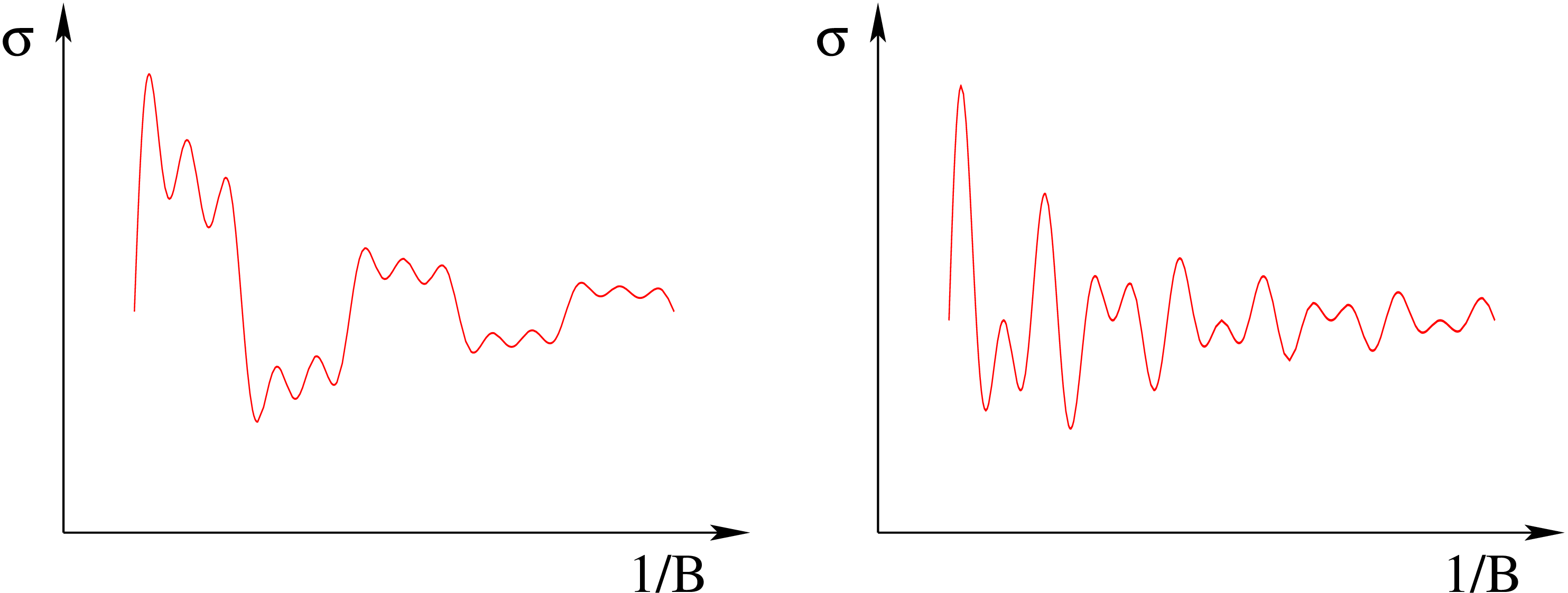}
\end{center}
\caption{The schematic sharp change of the experimental
picture of oscillations of the surface conductivity after 
crossing the boundary of a Stability Zone. It can be seen 
that the change in the oscillatory behavior is represented 
by the disappearance of one of the main oscillation terms 
from a finite sum of such terms.}
\label{CyclRes3}
\end{figure}

 Let us specially note here that both the vanishing of one
of the main oscillating terms in the conductivity and 
increasing of its period $\, T_{i_{0}} \, $ have very sharp 
character near the boundary of a Stability Zone. Indeed, 
the main oscillating terms in conductivity are given 
by the contributions of very narrow ``bands'' of closed 
trajectories near the extremal trajectories, which 
do not change noticeably up to the boundary of the Zone
$\, \Omega_{\alpha} \, $. In the same way, a noticeable
increasing of the period $\, T_{i_{0}} \, $ is caused by
a close approach of the singular trajectories to the extremal
trajectory on a cylinder of closed trajectories 
(Fig. \ref{Cylind}). It can be also shown, that this effect
starts to manifest itself in a pretty narrow region near
the boundary of $\, \Omega_{\alpha} \, $. As a result,
the intermediate picture represented at Fig. \ref{CyclRes1}
and Fig. \ref{CyclRes2} can in fact be practically 
unobservable in the experimental study of the cyclotron 
resonance phenomenon, such that we will most probably 
observe actually a more sharp transition 
(Fig. \ref{CyclRes3}) after crossing the 
boundary of a Stability Zone.

 We can see then that, despite the fact that the long closed 
trajectories arising outside a Stability Zone are hardly 
distinguishable from the open trajectories from experimental 
point of view, the cyclotron resonance phenomenon provides 
us a good tool for exact experimental determination of the 
boundary of a mathematical Stability Zone. Let us note also 
that the long closed trajectories arising near the boundary 
of a mathematical Stability Zone correspond to very large 
periods of electron motion along the orbit which can exceed 
the free electron motion time, so, they do not give in this 
case any visible contribution in the cyclotron resonance 
picture in experiment.

 Let us discuss now briefly quantum oscillations phenomena
in normal metals among which the oscillations of the
magnetic susceptibility (De Haas - Van Alphen effect) and
of the conductivity (Shubnikov - De Haas effect) with the value
of $\, 1/B \, $ are the most often mentioned. 

 The common reason for the quantum oscillation phenomena in
normal metals is the quantization of the electron motion 
along the closed trajectories of system (\ref{MFSyst}) in
the plane orthogonal to $\, {\bf B} \, $. In the quasiclassical
approximation we should put now that the closed trajectories
of system (\ref{MFSyst}) should be selected in accordance
with the quantization rule
$$S (\epsilon, p_{z}) \,\,\, = \,\,\,
{2 \pi e \hbar B \over c} \, \left( n + {1 \over 2} \right)
\,\,\, ,  \quad  n \gg 1 \,\,\, , $$
where $\, S (\epsilon, p_{z}) \, $ is the area bounded by a
closed trajectory of system (\ref{MFSyst}) in the plane
orthogonal to $\, {\bf B} \, $ in the ${\bf p}$ - space.
 
 The quantum oscillations of physical quantities measured in 
metals are connected then with the crossing of the Fermi level
by the corresponding quantized energy levels which defines
us the period of the corresponding oscillations, brought by a
fixed closed trajectory on the Fermi surface
$$\Delta \, {1 \over B} \,\,\, = \,\,\,
{2 \pi e \hbar \over c \, S (\epsilon_{F}, p_{z})} $$

 Like in the case of the cyclotron resonance the main terms
in the oscillating behavior are brought by the extremal closed
trajectories, which are defined now by the condition
$\, d S / d p_{z} \, = \, 0 \, $. As in the previous
situation, we can expect here the presence of a finite number
of trajectories of this kind on every cylinder, connecting
integral planes, if $\, {\bf B}/B \, \in \, \Omega_{\alpha} \, $.
Thus, the oscillating behavior of the physical quantities
should be mainly represented here by several main oscillating terms,
having different periods
$$\left( \Delta \, {1 \over B} \right)_{i} \,\,\, = \,\,\,
{2 \pi e \hbar \over c \, S^{i}_{extr}} $$
in the variable $\, 1/B \, $.

 For not extremely complicated Fermi surfaces we can expect again
that every cylinder of closed trajectories of system (\ref{MFSyst})
contains just one trajectory corresponding to an extremal area in 
the ${\bf p}$ - space. Besides that, for the Fermi surfaces of
not very high genus we can expect that every cylinder of closed
trajectories is invariant under the transformation
$\, {\bf p} \rightarrow - {\bf p} \, $, so, every extremal 
trajectory on this cylinder represents in fact its central
cross-section by the plane orthogonal to $\, {\bf B} \, $.
Let us say, however, that for very complicated Fermi surfaces
the assumptions above are not necessarily fulfilled. In particular,
in the situation when we have pairs of cylinders, transforming
into each other under the transformation
$\, {\bf p} \rightarrow - {\bf p} \, $, the extremal trajectories
on them can be located near the bases of the cylinders.  
In general, the details pointed above do not change noticeably
the scheme of using quantum oscillations phenomena for the 
determination of the exact boundary of Stability Zones for
magneto-conductivity in normal metals.  

 Like in the case of the cyclotron resonance, we should observe
here a ``quick change'' in the picture of oscillations of
physical quantities after crossing the boundary of 
a Stability Zone, which is caused by the disappearance
of one (or more) of the cylinders of closed trajectories
of system (\ref{MFSyst}). Also in this case the changes are 
sharply expressed at the boundary of $\, \Omega_{\alpha} \, $ 
since the corresponding main term in the oscillation picture 
is brought by an extremal trajectory,
which remains almost the same up to the boundary of
$\, \Omega_{\alpha} \, $ and disappears abruptly after
crossing the boundary. The changes in the oscillation picture
have the form, similar to that observed in the cyclotron 
resonance, showing the disappearance of one of the main
oscillation terms (Fig. \ref{CyclRes3}). Let us just note, 
that in this case we should not see at all an increasing of 
the corresponding period of oscillations near the boundary 
of $\, \Omega_{\alpha} \, $, since it is defined now by 
the area restricted by an extremal trajectory 
(and not cyclotron frequency), which does not change 
much near the boundary of a Stability Zone.

 At last, let us make one more additional remark. As we saw above,
the description of the quantum oscillations in normal metals is
connected with the area bounded by an extremal trajectory in the 
plane orthogonal to $\, {\bf B} \, $. The extremal trajectories 
in the theory of the cyclotron resonance are defined by the
extremal values of the cyclotron frequency, which is actually
connected with the value 
$\, \partial S (\epsilon, p_{z}) / \partial \epsilon \, $
according to the formulae
$$\omega_{B} \,\,\, = \,\,\, {e B \over m^{*} c} \quad ,  
\quad  \quad  m^{*} \,\,\, = \,\,\, {1 \over 2 \pi} \,
{\partial S (\epsilon, p_{z}) \over \partial \epsilon} $$

 At the same time, the value 
$\, \partial S (\epsilon, p_{z}) / \partial \epsilon \, $
can be measured also in the experimental study of the quantum
oscillations, using the investigation of their temperature
dependence (\cite{etm,Kittel,Abrikosov,Ziman}). We would like 
to point out here, that the values of
$\, \partial S (\epsilon, p_{z}) / \partial \epsilon \, $,
measured by these two different ways should coincide in fact
for $\, {\bf B}/B \, \in \, \Omega_{\alpha} \, $ for most
of the real metals having not extremely complicated
Fermi surfaces. The reason of this coincidence is given here
by the fact, that both the cyclotron resonance phenomenon
and the quantum oscillations phenomena are connected with the
same extremal trajectories, given by the central cross-sections
of the cylinders of closed trajectories in the topological
representation of the Fermi surface according to
Fig. \ref{ComplFermiSurf}. At the same time, for metals
with extremely complicated Fermi surfaces, this circumstance 
may not take place in general, so, the values of
$\, \partial S (\epsilon, p_{z}) / \partial \epsilon \, $,
measured in these two different ways, can be different,
since they are connected now with different extremal
trajectories of system (\ref{MFSyst}). It's not difficult
to see, that our last remark has no direct relationship
to the determination of the boundaries of the Stability Zones
in metals, however, it can play a role in more detailed
investigation of the oscillation phenomena for
$\, {\bf B}/B \, \in \, \Omega_{\alpha} \, $.

\section{Conclusions.}
\setcounter{equation}{0}

 We consider the problem of the exact determination of the
boundary of a Stability Zone for magneto-conductivity 
of normal metals in the space of directions of $\, {\bf B} \, $. 
As can be shown, this problem is actually rather nontrivial from 
experimental point of view due to a substantial difference
between the exact mathematical Stability Zones and the
``experimentally observable Stability Zones'' in the direct
conductivity measurements. It can be shown, however, that
the experimental detection of the exact boundary of a 
Stability Zone can be effectively implemented with the
aid of the well-known oscillation phenomena such as the
cyclotron resonance or quantum oscillations in normal metals.
Thus, the experimental study of oscillation phenomena in 
sufficiently strong magnetic fields reveals a sharp change 
in the picture of the oscillatory behavior of physical 
quantities after crossing the boundary of the mathematical 
Stability Zone as functions of the value of $B$ for a given 
direction of the magnetic field. This abrupt change in the 
oscillatory behavior can be described in the general case 
as the disappearance of one of the principal terms
from the total contribution, represented by a final sum 
of such terms (Fig. \ref{CyclRes3}). So, a detailed study 
of the oscillatory behavior of physical quantities at 
different directions of ${\bf B}$ allows us actually 
to determine the precise boundaries of the mathematical 
Stability Zones for the magneto-conductivity of metals. 
The results of the paper are based on the topological 
description of the structure of the Fermi surface in the 
case of presence of stable open quasiclassical electron 
trajectories on it.


\begin{thebibliography}{99}

\bibitem{lifazkag} I.M.Lifshitz, M.Ya.Azbel, M.I.Kaganov.
The Theory of Galvanomagnetic Effects in Metals.,
{\it Sov. Phys. JETP} {\bf 4}:1 (1957), 41.

\bibitem{lifpes1} I.M. Lifshitz, V.G. Peschansky.,
Galvanomagnetic characteristics of metals with open Fermi surfaces.,
{\it Sov. Phys. JETP} {\bf 8}:5 (1959), 875.

\bibitem{lifpes2} I.M. Lifshitz, V.G. Peschansky.,
Galvanomagnetic characteristics of metals with open Fermi surfaces. II.,
{\it Sov. Phys. JETP} {\bf 11}:1 (1960), 137.

\bibitem{lifkag1} I.M. Lifshitz, M.I. Kaganov.,
Some problems of the electron theory of metals I.
Classical and quantum mechanics of electrons in metals.,
{\it Sov. Phys. Usp.} {\bf 2}:6 (1960), 831-835.

\bibitem{lifkag2} I.M. Lifshitz, M.I. Kaganov.,
Some problems of the electron theory of metals II.
Statistical mechanics and thermodynamics of electrons in metals.,
{\it Sov. Phys. Usp.} {\bf 5}:6 (1963), 878-907.

\bibitem{lifkag3} I.M. Lifshitz, M.I. Kaganov.,
Some problems of the electron theory of metals III.
Kinetic properties of electrons in metals.,
{\it Sov. Phys. Usp.} {\bf 8}:6 (1966), 805-851.

\bibitem{etm} I.M. Lifshitz, M.Ya. Azbel, M.I. Kaganov.,
Electron Theory of Metals. Moscow, Nauka, 1971.
Translated: New York: Consultants Bureau, 1973.

\bibitem{KaganovPeschansky} M.I. Kaganov, V.G. Peschansky.,
Galvano-magnetic phenomena today and forty years ago.,
{\it Physics Reports} {\bf 372} (2002), 445-487.

\bibitem{MultValAnMorseTheory} S.P. Novikov.,
The Hamiltonian formalism and a many-valued analogue of
Morse theory., {\it Russian Math. Surveys} {\bf 37} (5) (1982), 1-56.

\bibitem{zorich1} A.V. Zorich.,
A problem of Novikov on the semiclassical motion of an electron in a 
uniform almost rational magnetic field.,
{\it Russian Math. Surveys} {\bf 39} (5) (1984), 287-288.

\bibitem{dynn1992} I.A. Dynnikov.,
Proof of S.P. Novikov's conjecture for the case of small perturbations 
of rational magnetic fields.,
{\it Russian Math. Surveys} {\bf 47} (3) (1992), 172-173.

\bibitem{Tsarev} S.P. Tsarev.
Private communication. (1992-93).

\bibitem{dynn1} I.A. Dynnikov.,
Proof S.P. Novikov's conjecture on
the semiclassical motion of an electron.,
{\it Math. Notes} {\bf 53}:5 (1993), 495-501.

\bibitem{dynn2} I.A. Dynnikov.,
Semiclassical motion of the electron. A proof of the Novikov conjecture
in general position and counterexamples., Solitons, geometry, and
topology: on the crossroad, Amer. Math. Soc. Transl. Ser. 2, 179,
Amer. Math. Soc., Providence, RI, 1997, 45-73.

\bibitem{PismaZhETF}
S.P. Novikov, A.Y. Maltsev.,
Topological quantum characteristics observed in the investigation of the
conductivity in normal metals., {\it JETP Letters}
{\bf 63} (10) (1996), 855-860.

\bibitem{dynn3} I.A. Dynnikov.,
The geometry of stability regions in Novikov's problem on the
semiclassical motion of an electron.,
{\it Russian Math. Surveys} {\bf 54}:1 (1999), 21-59.

\bibitem{UFN} S.P. Novikov, A.Y. Maltsev.,
Topological phenomena in normal metals.,
{\it Physics-Uspekhi} {\bf 41}:3 (1998), 231-239.

\bibitem{BullBrazMathSoc}
A.Ya. Maltsev, S.P. Novikov.,
Quasiperiodic functions and Dynamical Systems
in Quantum Solid State Physics.,
{\it Bulletin of Braz. Math. Society}, New Series {\bf 34}:1 (2003),
171-210.

\bibitem{JournStatPhys}
A.Ya. Maltsev, S.P. Novikov.,
Dynamical Systems, Topology and Conductivity in Normal Metals in
strong magnetic fields.,
{\it Journal of Statistical Physics} {\bf 115}:(1-2) (2004), 31-46.

\bibitem{DeLeoPhysB} R. De Leo.,
First-principles generation of stereographic maps for high-field 
magnetoresistance in normal metals: An application to Au and Ag.,
{\it Physica B: Condensed Matter} {\bf 362} (1–4) (2005), 62–75.

\bibitem{zorich2} A.V. Zorich.,
Proc. ``Geometric Study of Foliations''.,
(Tokyo, November 1993) / ed. T.Mizutani et al.
Singapore: World Scientific, 479-498 (1994).

\bibitem{ZhETF1997} A.Y. Maltsev.,
Anomalous behavior of the electrical conductivity tensor
in strong magnetic fields., {\it JETP} {\bf 85} (5) (1997), 934-942.

\bibitem{DeLeo1} R. De Leo.,
Existence and measure of ergodic leaves in Novikov’s problem
on the semiclassical motion of an electron.,
{\it Russian Math. Surveys} {\bf 55}:1 (2000), 166-168.

\bibitem{DeLeo2} R. De Leo.,
Characterization of the set of ``ergodic directions'' in Novikov's
problem of quasi-electron orbits in normal metals.,
{\it Russian Math. Surveys} {\bf 58}:5 (2003), 1042-1043.

\bibitem{DeLeo3} R. De Leo.,
Topology of plane sections of periodic polyhedra with an application
to the Truncated Octahedron.,
{\it Experimental Mathematics} {\bf 15}:1 (2006), 109-124.

\bibitem{DeLeoDynnikov1} R. De Leo, I.A. Dynnikov.,
An example of a fractal set of plane directions having chaotic
intersections with a fixed 3-periodic surface.,
{\it Russian Math. Surveys} {\bf 62}:5 (2007), 990–992.

\bibitem{DeLeoDynnikov2} R. De Leo, I.A. Dynnikov.,
Geometry of plane sections of the infinite regular skew polyhedron
$\{ 4, \, 6 \, | \, 4 \}$.,
{\it Geom. Dedicata} {\bf 138}:1 (2009), 51-67.

\bibitem{Skripchenko1} A. Skripchenko.,
Symmetric interval identification systems of order three.,
{\it Discrete Contin. Dyn. Sys.} {\bf 32}:2 (2012), 643-656.

\bibitem{Skripchenko2} A. Skripchenko.,
On connectedness of chaotic sections of some 3-periodic surfaces.,
{\it Ann. Glob. Anal. Geom.} {\bf 43} (2013), 253-271.

\bibitem{DynnSkrip} I. Dynnikov, A. Skripchenko.,
On typical leaves of a measured foliated 2-complex of thin type.,
Topology, Geometry, Integrable Systems, and Mathematical Physics:
Novikov's Seminar 2012-2014, Advances in the Mathematical Sciences.,
Amer. Math. Soc. Transl. Ser. 2, 234, eds. V.M. Buchstaber,
B.A. Dubrovin, I.M. Krichever, Amer. Math. Soc., Providence,
RI, 2014, 173-200, arXiv: 1309.4884

\bibitem{GurzhyKop} R.N. Gurzhy, A.I Kopeliovich.,
Low-temperature electric conductivity of pure metals.,
In the book: Conductivity electrons., Red. M.I. Kaganov, 
V.S. Edelman., Moscow, Nauka, 1985 (in Russian).

\bibitem{JETP2017} A.Ya. Maltsev.,
On the Analytical Properties of the Magneto-Conductivity
in the Case of Presence of Stable Open Electron Trajectories
on a Complex Fermi Surface., 
{\it Journal of Experimental and Theoretical Physics} {\bf 124} 
(5) (2017), 805-831.

\bibitem{Kittel} C. Kittel.,
Quantum Theory of Solids., Wiley, 1963.

\bibitem{Abrikosov} A.A. Abrikosov.,
Fundamentals of the Theory of Metals.,
Elsevier Science \& Technology, Oxford, United Kingdom, 1988.

\bibitem{Ziman} J.M. Ziman.,
Principles of the Theory of Solids., Cambridge University 
Press 1972.




\end{thebibliography}
\end{document}